\input{psfig}
\documentclass[referee]{aa}

\usepackage{psfig}

\thesaurus{12.03.4; 12.12.1; 11.06.1}
\title{Non-local bias and the problem of large-scale power in the {\it 
Standard} Cold Dark Matter model}
\author{A. Del Popolo\inst{1,2} \and Y. Takahashi\inst{3} \and H. Kiuchi\inst{3} \and M. Gambera\inst{1,4}}
\titlerunning{Non local bias and the SCDM model}
\authorrunning{Del Popolo et al.}
\date{}
\offprints{A. Del Popolo, E-mail:antonino@astrct.ct.astro.it OR  M. Gambera  E-mail:mga@sunct.ct.astro.it}
\institute{$^1$ Istituto di Astronomia dell'Universit\`a di Catania, 
Viale A.Doria, 6 - I 95125 Catania, ITALY \\
$^2$ Dipartimento di Matematica, Universit\`{a} Statale di Bergamo,
Piazza Rosate, 2 - I 24129 Bergamo, ITALY \\
$^3$ Communications Reasearch Laboratory, 4-2-1 Nukui-kita, Koganei, Tokyo 184-8795, JAPAN \\
$^4$ Osservatorio Astrofisico di Catania and CNR-GNA, 
Viale A.Doria, 6 - I 95125 Catania, ITALY 
}
\begin{document}
\maketitle
\begin{abstract}
We study the effect of non-radial motions, originating from the
gravitational interaction of the quadrupole moment of a protogalaxy with
the tidal field of the matter of the neighboring protostructures, on the
angular correlation function of galaxies. We calculate the angular
correlation function using a {\it Standard} Cold Dark Matter (hereafter 
SCDM) model ($\Omega=1$, h=0.5, $n=1$) and we compare it with the angular
correlation function of the APM galaxy survey (Maddox et al. 1990; Maddox et al. 1996).
We find
that taking account of non-radial motions in the calculation of the angular
correlation function gives a better agreement of the theoretical
prediction of the SCDM model to the observed estimates of large-scale power
in the galaxy distribution. 
\end{abstract}

\section{Introduction}

The galaxy two-point correlation function $\xi_g (r)$ is a powerful
discriminant between distinct models of structure formation in the
universe. On scales $\geq 10h^{-1}{\rm Mpc}$ correlations between galaxies are
weak, namely $\xi_g (r)<<1$, so one may reasonably expect that $\xi_g (r)$ can
be related to the fluctuations in the early universe by linear perturbation
theory. The role of the two-points correlation function is particularly
important in models that predict fluctuations that obey Gaussian statistics
(see Bardeen et al. 1986) because in this case the field is completely
specified, in a statistical sense, by a single function: the power spectrum $%
P(k)$ or by its Fourier transform, the autocorrelation function, $\xi (r)$.
This means that a knowledge of $\xi_g (r)$ on large scales would give a
powerful constraint on models of the early Universe, and if the fluctuations
are Gaussian, we would obtain a complete description of large-scale
structure. Analyses of galaxy surveys (APM, QDOT) have shown that excess
correlations are found on scales larger than $10h^{-1}$ {\rm Mpc} (Maddox et. al 1990;
Efstathiou et al. 1990b; Saunders et al. 1991; Maddox et al. 1996) when observations are compared with the predictions of the
standard CDM model and also radio galaxies are also strongly clustered on
large scales (Peacock 1991; Peacock \& Nicholson 1991). These
data have hence been widely interpreted as ruling out SCDM model and new
alternative models have been introduced in an effort to solve this and other
problems of the model. \\
Several authors (Peebles 1984; Efstathiou et al.
1990a; Turner 1991) have lowered the matter density under the critical value
($\Omega _m<1$) and have added a cosmological constant in order to retain a
flat Universe ($\Omega _m+\Omega _\Lambda =1$). Mixed dark matter models
(MDM) (Shafi \& Stecker 1984; Valdarnini \& Bonometto 1985; Schaefer et al.
1989; Holtzman 1989; Schaefer 1991; Schaefer \& Shafi 1993; Holtzman \&
Primack 1993) increase the large-scale power because neutrinos
free-streaming damps the power on small scales. Alternatively, changing the
primeval spectrum solves several problems of CDM (Cen et al. 1992).
Finally it is possible to assume that the threshold for galaxy formation is
not spatially invariant but weakly modulated ($2\% \div 3\%$ on scales $%
r>10h^{-1}{\rm Mpc}$) by large scale density fluctuations, with the result that
the clustering on large-scale is significantly increased (Bower et al.
1993). This last alternative is part of those models attempting to rescue
SCDM by invoking a form of bias that has different effects on clustering on
different scales (Babul \& White 1991; Bower et al. 1993). In fact as
previously reported, large-scale clustering studies such as APM (Maddox et
al. 1990; Maddox et al. 1996)
and QDOT\ (Efstathiou et al. 1990b; Saunders et al. 1991; Peacock
1991) suggest a clustering amplitude which is larger than one would expect
on the basis of the SCDM model. Moreover the level of temperature
fluctuations seen by COBE are consistent with no large-scale bias, but when
the CDM model is normalized to COBE results it has problems in
accounting for 
small-scale structure. An obvious way to rescue the model, then, is a
scale-dependent bias which can modify the slope of the correlation function
so as to make it decay less steeply than the mass autocovariance function
on large scales. As shown by Coles (1993) for Gaussian fields, a change of
slope of this kind can be achieved by non-local biasing effects such as
cooperative galaxy formation (Bower et al. 1993). In this last model
Bower et al.
(1993) adopt the assumption that the threshold
level, $\delta _c$, depends on the mean mass density in the domain of
influence, rather than being spatially invariant. Galaxy formation is assumed
to occur according to the prescriptions of the standard biased galaxy
formation theory (Kaiser 1984; Bardeen et al. 1986) but is enhanced by the
presence of nearby galaxies. This approach is able to produce enough
additional clustering to fit the $\xi_g (r)$ of the APM galaxy survey. The
main problem of this and of any theory of galaxy formation involving a bias
of any kind is that they are not acceptable until the physical mechanisms
producing the bias are elucidated. In some recent papers (Del Popolo \&
Gambera 1998a,b) we introduced a model that
is
able to reduce several of the SCDM model problems and also includes
(differently from Bower et al. 1993) a clear
explanation for the physical mechanisms that produce the bias. \\
As shown by
Barrow \& Silk (1981) and Szalay \& Silk (1983) the gravitational
interaction of the irregular mass distribution of a test
proto-structure with the
neighbouring ones gives rise to non-radial motions, within the
test proto-structure, which are expected to slow the rate of growth of the density
contrast and to delay or suppress the collapse. According to Davis \&
Peebles (1977), Villumsen \& Davis (1986) and Peebles (1990) the kinetic
energy of the resulting non-radial motions at the epoch of maximum expansion
increases so much as to oppose the recollapse of the proto-structure.
As shown by Del Popolo \& Gambera (1998a,b), 
within high-density environments, such as rich clusters of galaxies,
non-radial motions slow down the collapse of low-$\nu $ peaks thus producing an
observable variation in the time of collapse of the shell and, as a
consequence, a reduction in the mass bound to the collapsed perturbation.
Moreover, the delay of the collapse produces a tendency for less dense
regions to accrete less mass, with respect to a classical spherical model,
inducing a {\it biasing} of over-dense regions toward higher mass.
Non-radial motions change the energetics of the collapse model by
introducing another potential energy term in the equation of collapse,
leading to a change
of the {\it turn around} epoch, $t_{m}$, and consequently the critical
threshold, $\delta_{c}$, for collapse. 
The change of $\delta_c$ is in the
same sense as that described by Bower et al. (1993). \\
In this paper we apply the
quoted model to galaxies showing how non-radial motions modify
the galaxies correlations. The plan of the paper is the following:
in Sect. ~2 we
introduce the model; in Sect. ~3 show the results and in
Sect.~4 we draw our conclusions.

\section{The model}

In the standard high-peak galaxy model, galaxies form from mass located near
high peaks of the linear density field. The density contrast
at early times , $\delta ({\bf x})=\frac{\rho ({\bf x)-}\rho _b}{\rho _b}$
is assumed to be 
Gaussian, and the field $\delta ({\bf x})$ is
smoothed by convolving it with
a spherical symmetric window function $W(r,R_g)$, where the characterstic
scale $R_g$ is chosen so that the enclosed mass matches the halo of a bright
galaxy. Galaxy formation sites are identified with peaks rising above a
threshold, $\delta >$ $\delta _c$. The value of $\delta _c$ is quite
dependent on the choice of smoothing window used to obtain the dispersion
(Lacey \& Cole 1994). Using a top-hat window function $\delta _c=1.7\pm 0.1$,
while for a Gaussian window the threshold is significantly lower. In
non-spherical situation things are more complicated (Monaco 1995). In any case in
the standard biased galaxy formation the threshold is not scale-dependent
and is taken to be universal. \\
Several studies have shown that there is no
convincing justification for this choice (Cen \& Ostriker 1992; Bower et al.
1993; Coles 1993; Del Popolo \& Gambera 1998a,b,c;
Kauffmann et al. 1998; Willmer et al. 1998; Governato et al. 1998; Peacock
1998). \\
Some authors (see Barrow \& Silk 1981; Szalay \& Silk 1983 and Peebles 1990)
have proposed that non-radial motions would be expected within a developing
proto-galaxy due to the tidal interaction of the irregular mass distribution
around them, typical of hierarchical clustering models, with the
neighbouring proto-galaxies. The kinetic energy of these non-radial motions
prevents the collapse of the proto-structure, enabling the same to reach
statistical equilibrium before the final collapse (the so-called
previrialization conjecture by Davis \& Peebles 1977, Peebles 1990). In
other words one expects that non-radial motions change the characteristics
of the collapse and in particular the {\it turn around} epoch, $t_{m}$, and
consequently the critical threshold, $\delta_{c}$, for collapse.

\begin{figure}
\psfig{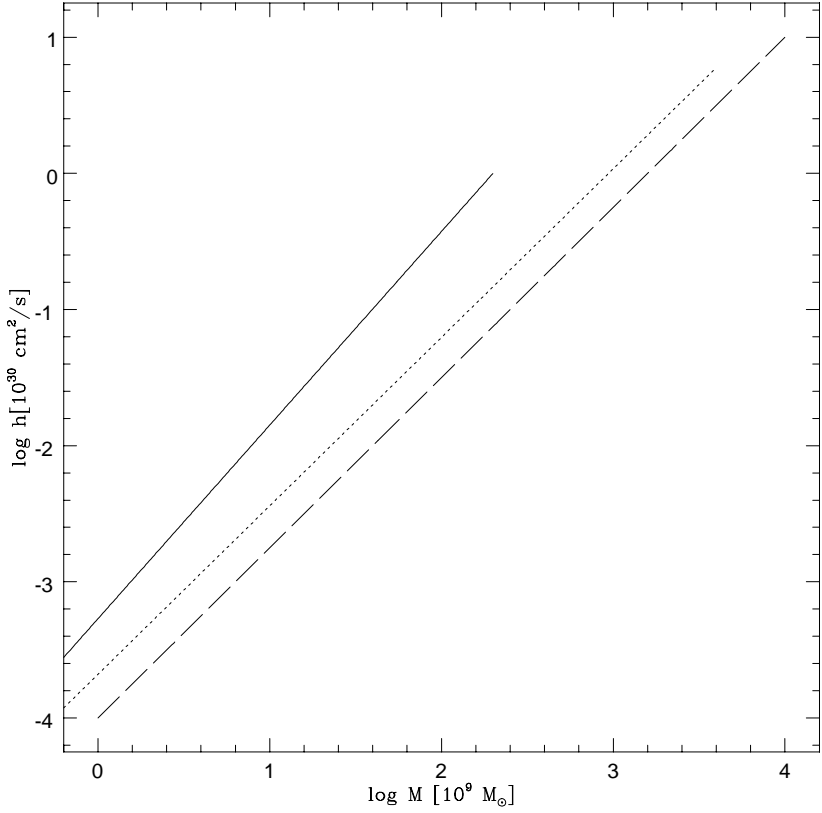}
\caption[]{The specific angular momentum for
three values of the parameter $\nu$ ($\nu=2$ solid line, $\nu=3$
dotted line, $\nu=4$ dashed line).}
\label{Fig. 1}
\end{figure}

As shown by Del Popolo \& Gambera (1998a,b,c), if non-radial motions are taken
into account, the threshold $\delta_c$ is not constant but is function of
mass, $M$, (Del Popolo \& Gambera 1998a,b): 
\begin{equation}
\delta _c(\nu )=\delta _{co}\left[ 1+\frac{8G^2}{\Omega _o^3H_0^6r_i^{10}%
\overline{\delta} (1+\overline{\delta} )^3}\int_{a_{min}}^{a_{\max }}\frac{%
L^2 \cdot da}{a^3}\right]  \label{eq:ma8}
\end{equation}
where $\delta _{co}=1.68$ is the critical threshold for a spherical model, $%
r_i$ is the initial radius, $L$ the angular momentum, $H_0$ and $\Omega_0$
the Hubble constant and the density parameter at the current epoch, respectively, 
$a$ the expansion parameter and $\overline{\delta}$ the mean fractional
density excess inside a shell of given radius. The mass dependence of the
threshold parameter, $\delta_{c}(\nu)$, and the total specific angular
momentum, $h(r,\nu )=L(r,\nu)/M_{sh}$, acquired during expansion, was
obtained in the same way as described in Del Popolo \& Gambera (1998b) and is
displayed in Fig. ~\ref{Fig. 1}. \\

In order to find the galaxy correlation function, we combine Kaiser`s (1984)
analysis with the theory of gravitational clustering
(Press \& Schecter 1974). In this last theory non-linear clumps are
identified as overdensities in the filtered linear density field. When these
overdensities exceed a critical threshold, $\delta_c$, they will be
incorporated in a non-linear object of mass $M \propto R^3_g$ or greater.
Since the linear density field is assumed to be Gaussian, the
probability that on scale $M$ one would find a density contrast
between $\delta $ and $\delta +d\delta $ would be:
\begin{equation}
p(\delta )d\delta =\frac 1{\sqrt{2\pi }\sigma (M)}\exp \left[-\frac{\delta ^2%
}{2\sigma (M)^2}\right]d\delta
\end{equation}
and the Press-Schecter ansatz leads to the following fraction of mass incorporated in
objects of mass $>M$ (or 
the probability that an object of mass $M$ has turned around at any time in
the past):
\begin{equation}
P_M=\int_{\delta _c}^\infty p(\delta )d\delta =\frac 12 erfc(\frac{\delta _c%
}{\sqrt{2}\sigma })
\end{equation}
And in a similar way, the probability density for finding  
simultaneously, on scale $M$, the density contrast $\delta _1=\delta({\bf x}_1)$
and $\delta _2=\delta({\bf x}_2)$ of two field points 
separated by $r=|{\bf x}_1-{\bf x}_2|$ is:
\begin{eqnarray}
p(\delta _1,\delta _2,\xi/\sigma^2) &=& \frac 1{2\pi
\sqrt{\xi^2(0)-\xi ^2}}  \cdot \nonumber \\
& \times& \exp \left[-\frac{\xi(0) \delta^2 _1 
+\xi(0) \delta^2 _2-2 \xi \delta _1 \delta _2}{2(\xi^2(0)-
\xi ^2)}%
\right]
\end{eqnarray}
where 
\begin{equation}
\xi (r)=\frac 1{2\pi ^2}\int_0^\infty P(k)k^2\frac{\sin (kr)}{kr}\exp
(-k^2R_g^2)
\end{equation}
is the mass autocorrelation function of the primeval density distribution, $%
\delta _1$ and $\delta _2$ are used to denote the density contrast at the
positions ${\bf x}_1$ and ${\bf x}_2$ of the two field points and
$\sigma$ denotes the r.m.s. of $\delta$.
The power spectrum used has the
form given by Bardeen et al. (1986): 
\begin{eqnarray}
P(k) &=& A k^{-1}[\ln \left( 1+4.164k\right)]^2 \cdot (192.9+1340k+  \nonumber \\
& + & 1.599\cdot 10^5k^2+1.78\cdot 10^5k^3+3.995\cdot 10^6k^4)^{-1/2}
\label{eq:ma5}
\end{eqnarray}
and $A$ is the normalizing constant, which gives the amplitude of the
power spectrum. Similarly to Bower et al. (1993),
since our model, like the original high-peak model, calculates
$\xi_{\rm g}$ from $\xi/\sigma^2$,the amplitude of the power spectrum
drops out of our analysis.
The probability of finding two objects
of masses $M$ separated by $r$ that have turned around in
any time between the actual time and $t=0$ is: 
\begin{eqnarray}
P_{M M} = \int_{\delta _c}^\infty \int_{\delta _c}^\infty p(\delta
_1,\delta _2,\xi/\sigma^2)d\delta _1d\delta _2 &=&  \nonumber \\
\frac 1{2\sigma \sqrt{2\pi }} \int_{\delta _c}^\infty \exp (-\frac{\delta
_1^2}{2\sigma^2})erfc\left[ \frac{\nu-
\frac{\xi}{\xi(0)} \frac{%
\delta _1}{\sigma }}{\sqrt{2\left( 1-
\xi ^2/\xi^2(0)\right) }}\right] d\delta _1
\end{eqnarray}
Following Kaiser (1984), the correlation
function of peaks above a given threshold,
on scales larger than $R_g$, can be approximated by that of
points above the same threshold. According to Kaiser's (1984) definition we
have:
\begin{equation}
\xi _g(r)=\frac{P_{M M}}{P^2_{M}}-1  \label{eq:corr}
\end{equation}
giving the fractional excess probability that two points at separation $r$
are both above the threshold. Eq. (\ref{eq:corr}) shows that the $\xi _g(r)$
is a function of $\xi (r)/\sigma ^2$ and of $\delta _c$. In the limit $\delta
_c\rightarrow \infty $, $\xi \rightarrow 0$ we have:
\begin{equation}
\xi _g(r)\simeq \left( \frac{\delta _c}\sigma \right) ^2\xi (r)
\label{eq:core}
\end{equation}
(Kaiser 1984). This approximation is not so accurate on the scales
we are considering, we thus prefer
to evaluate $\xi_g(r)$ numerically after having reduced
the dimensionality of the integrals involved in the calculation,
as Bower et al. (1993) have done,
by means of:
\begin{eqnarray}
\xi _g(r)&=&\xi (r)\int_0^1\left[ \xi (0)^2-s^2\xi (r)^2\right] ^{-1/2}\exp
\left[ -\frac{\delta _c^2\xi (0)}{\xi (0)+s\xi (r)}\right] ds  \nonumber \\
&\times& \left[ \int_{\delta _c}^\infty \exp (-\frac{u^2}2)du\right] ^{-2}
\end{eqnarray}
In order to compare our model predictions of large-scale power in galaxy
distribution with its estimates from the APM survey (Maddox et al. 1990) we have
to calculate the angular two points autocorrelation function,
$w(\vartheta )$. This last is related to the spatial correlation function,
$\xi(r)$, through Limber's (1954) equation 
%
%
%
(see also Peebles 1980, Peacock 1991):
\begin{equation}
w(\vartheta )=\int_0^\infty y^4\phi ^2dy\int_{-\infty }^\infty \xi (\sqrt{%
x^2+y^2\theta ^2})dx/\left[ \int_0^\infty y^2\phi dy\right]  \label{eq:ang}
\end{equation}
where the luminosity function, $\phi (y)$, is that recommended by Maddox et
al. (1990) and where:
\begin{equation}
\phi (y)dy=\phi ^{*}y^\alpha \exp (-y)dy
\end{equation}
being
\begin{equation}
y=10^{0.4(M_b^{*}(z)-M)} 
\end{equation}
and
\begin{equation}
\phi ^{*}=1.3\times 10^{-2}h^3{\rm Mpc}^{-3}
\end{equation}
with
\[
\begin{array}{l}
M_b^{*}(z) = M_0^{*}+M_1^{*}z \\
\alpha (z) = \alpha _0+\alpha _1z \\
M_0^{*} = -19.8 \\
M_1^{*} = 1 \\
\alpha _0 = -1 \\
 \alpha _1 = -2 \\
\end{array}
\]

In order to calculated Eq. (\ref{eq:ang}) we need the spatial counterpart $%
\xi (r)$ for all $r$ . We have used an approach similar to that by Maddox et
al. (1990) and Bower et al. (1993), namely we calculated
correlation functions according to what we have previously described in this
section but on small scales ($r\leq 5.7h^{-1}{\rm Mpc}$)
we extrapolated our model correlation
functions by using
$\xi (r)=\left( \frac{5.7h^{-1}{\rm Mpc}}{r}\right) ^{1.7}$.

\section{Results and discussion}

\begin{figure}
\psfig{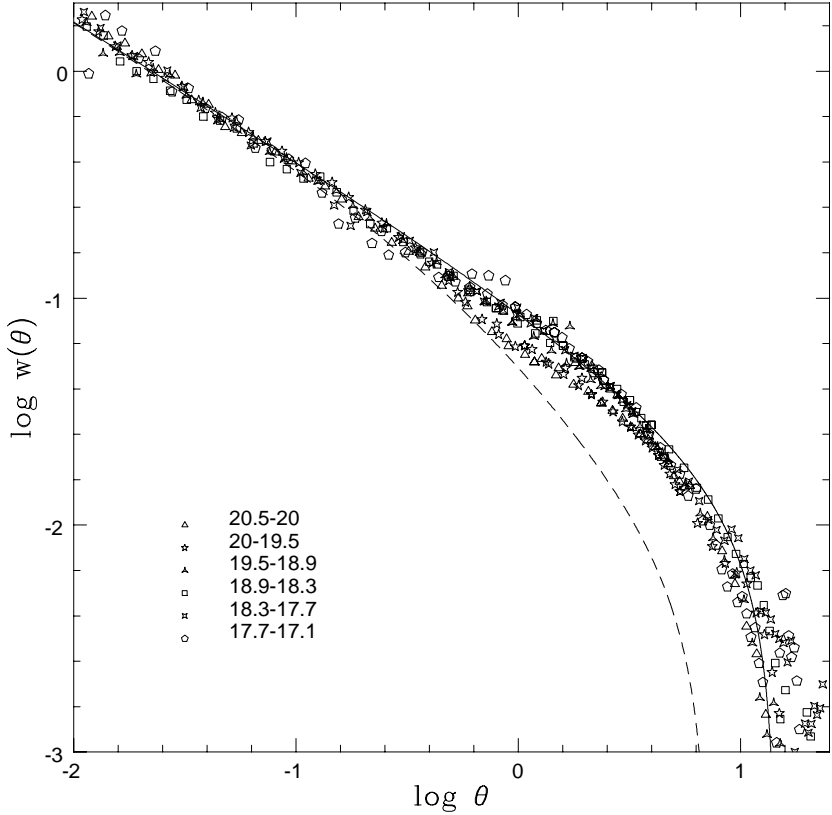}
\caption[]{Angular correlation function for galaxies for the APM survey
and in CDM models. The data (kindly provided by W. Sutherland)
represents estimates of $w(\theta)$ for six
disjoint magnitude slices in the range $17 \leq b_j \leq 20.5$
scaled to the magnitude limit of the Lick catalogue, $b_j=18.4$, published
in Maddox et al. 1996. The dashed line shows correlations in the
SCDM model while
the solid line shows correlations in our model.}
\label{Fig. 2}
\end{figure}

The result of our model is directly compared (see Fig. ~\ref{Fig. 2}) with the angular correlation
function estimate by Maddox et al. (1990) from the APM survey, in order to
find whether it can match observed estimates of large-scale power in
the galaxy distribution. The data
(kindly provided by W. Sutherland) plot the angular correlation function, $%
w(\theta )$, for six disjoint apparent magnitude slices between $17\leq
b_j\leq 20.5$, all scaled to the magnitude limit of the Lick catalogue
(Groth \& Peebles 1977), $b_j=18.4$.
At small angles the angular correlation function is a power law:
\begin{equation}
w(\theta)=B \theta^{1-\gamma}
\end{equation}
For $0.01^o < \theta < 1^o$ the values of $\gamma$ and $B$ are:

\[
\begin{array}{l}
\gamma=1.699 \pm 0.032\\
B=0.0284 \pm 0.0029\\
\end{array}
\]

\begin{figure}
\psfig{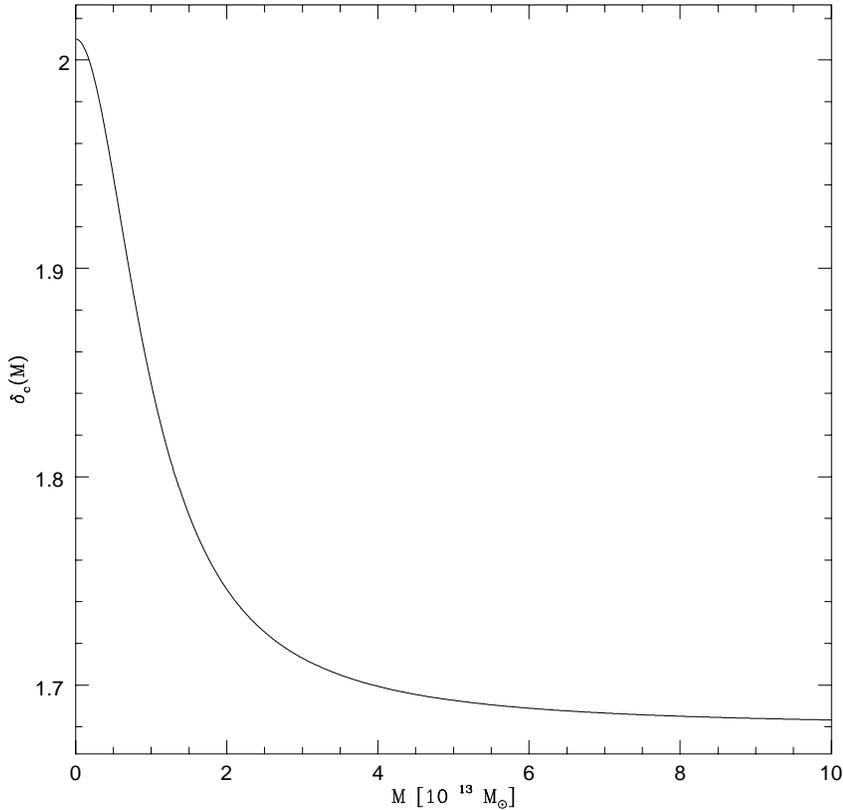}
\caption[]{The threshold $\delta_{c}$  as a function of the mass M, for a CDM
spectrum ($\Omega_0=1$, $h=1/2$),
taking account of non-radial motions.}
\label{Fig. 3}
\end{figure}

(Maddox et al. 1996). At larger angles the angular correlation 
function steepens and lies
below the extrapolation of the power law. At magnitude $b_j=20$ the steepening occurs
at $ \simeq 2^o$.\\ 
The dotted line shows the angular
correlation function calculated using the SCDM model, then assuming a uniform biasing
threshold. Fig. ~\ref{Fig. 2} clearly shows the well SCDM known problem of
lack of large-scale power, namely the two-point angular correlation
function, when fit to the observations on the 0.03-0.3 degree scale,
is significantly below the observations on scales greater than $1$
degree, when these are scaled to the depth of the Lick survey. This
provides strong evidence for large-scale power in the galaxy distribution
that cannot be reconciled with the SCDM model. The same conclusion is
achieved by analysis of three-dimensional data (Vogeley et al. 1992) and
from an independent redshift survey of a subset of the APM-Stromlo galaxies
(Loveday et al. 1992).
An important point to stress is that
this problem is normalization independent, it cannot be solved by changing the bias
strength,
(Maddox et al. 1990; Bartlett \&
Silk 1993; Ostriker 1993) because the theoretical $w(\theta )$ has the wrong
shape. \\
The solid line shows the angular correlation function obtained in our
model (CDM with $\Omega=1$, $h=0.5$ and taking account of non-radial motions).
Our $w(\vartheta )$ is less steep, at large angles,
than that expected from the SCDM model
and is in better agreement with observations.
The relatively enhanced power on large scales, with respect to
smaller scales,
is insensitive to the amplitude of the power spectrum:
the better agreement with
observations is due to the non uniform biasing threshold in our model
as given by Eq. (\ref{eq:ma8}). \\
Similarly to what is shown in
Del Popolo \& Gambera (1998a,b), Fig. ~\ref{Fig. 3} shows that 
the threshold, $\delta_c$, is a
decreasing function of the mass, $M$. This means that peaks in more dense
regions must have a lower value of the threshold, $\delta_c$, with respect
to those of under-dense regions, in order to form structure.
In fact, as clearly shown in Fig. ~\ref{Fig. 1},
the angular momentum acquired by a shell centered on a peak
in the CDM density distribution is anti-correlated with density: high-density
peaks acquire less angular momentum than low-density peaks
(Hoffman 1986; Ryden 1988).
A greater amount of
angular momentum acquired by low-density peaks
(with respect to the high-density ones)
implies that these peaks can more easily resist
gravitational collapse and consequently it is more difficult for them to form
structure.
This results in a tendency for less dense
regions to accrete less mass, with respect to a classical spherical model,
inducing a {\it biasing} of over-dense regions toward higher mass.
This also explains why the value of $\delta_c$,
that a peak must rise above in order to form a structure,
is larger for low-mass peaks than high
density ones.
The space dependence of the threshold implies also a scale dependence of the
bias parameter, $b$, because the two parameters are connected (see Borgani 1990;
Mo \& White 1996; Del Popolo \& Gambera 1998a,b,c). \\
Another way of describing the differences between
our model and the standard biased galaxy formation is the following.
In order to describe the distribution of objects we consider the biased field:
\begin{equation}
\rho_{\delta_c,R_g}(x)=t[\delta_R(x)-\delta_c]
\end{equation}
where $t(y)$, the threshold function, relates the biased field
$\rho_{\delta_c, R_g}(x)$ to the background field,
$\rho_{R_g}(x)$. The threshold
function, $0 \leq t(y) \leq 1$, gives the probability that a fluctuation of a given amplitude
turns out to be revealed as an object, 
while the threshold level $\delta_c$ is defined as the value of the density
contrast at which a fluctuation has a probability of $1/2$ of giving
rise to an object. The simplest case is that of the standard biased galaxy
formation in which
the selection
function, is a Heaviside function
$t(y)=\theta(y)$ and $\delta_c$ is the limiting
height of the selected fluctuations.
In this scheme, fluctuations below $\delta_c$ have zero
probability of developing an observable object and fluctuations above $\delta_c$
have zero probability not to develop an object.
As we showed in a previous paper (Del Popolo \& Gambera 1998a)
one of the effects of non-radial motions is that the threshold function
differs from a Heaviside function (sharp threshold),
(see Fig. 7 of Del Popolo \& Gambera 1998a).
This means that 
objects can also be formed from fluctuations
below $\delta_c$ and
that there is a non-zero probability for fluctuations above $\delta_c$ to
be sterile. \\
The model is
similar to the cooperative galaxy formation theory but now there is a simple
explanation for the mass dependence of
the threshold, $\delta_c$: it is due to non-radial motions. As  we said
previously
non-radial motions arise from the gravitational
interaction of the quadrupole moment of the system with the tidal field of
the matter of the neighbouring proto-galaxies (Barrow \& Silk 1981;
Szalay \& Silk 1983; Peebles 1990). The energy connected to these motions
enters the equation of spherical collapse thus changing the turnaround epoch
and $\delta_c$. Being Eq. (\ref{eq:core}) dependent on the threshold,
$\delta_c$, the galaxy correlation function is changed as well.
The final result is an increase of the galaxy correlations, $\xi_g$,
predicted on large scales.

\section{Conclusion}

The galaxy two-point correlation function $\xi_g (r)$ has a special place
amongst statistics of the galaxy distribution and is a powerful
discriminant between distinct models of structure formation in the
universe. Several studies (Maddox et al. 1990, Efstathiou et al. 1990;
Saunders et al. 1991; Loveday et al. 1992; Maddox et al. 1996)
have shown that the 
$\xi_g(r)$ obtained from an SCDM model, independently from normalization,
is difficult to reconcile with the observed $\xi_g(r)$ if the bias is scale
independent. In this paper we showed how a scale dependent bias, due to
non-radial motions, can reduce the problem of large-scale lack of power
in the SCDM model. We calculated the mass dependence of the
threshold parameter, $\delta_c$,
(due to non-radial motions) and then used it to find the two-points
correlation function following Press \& Schecter (1974) and Kaiser (1984).
This was used to find the angular correlation function, $w(\theta)$,
through Limber`s (1954) equation. The $w(\theta)$ found in this way was
compared with data of the APM survey
(Maddox et al. 1990; Maddox et al. 1996). We found a less steep
$w(\theta)$ in good agreement with estimates of large-scale power in the galaxy
distribution.

\begin{acknowledgements}
We would like to thank the referee R. Cen for helpful comments
which led to improve our paper. 
We are grateful to Prof. E. Recami and Prof. E. Spedicato for stimulating
discussions while this work was in process.
A. D.P. would like to thank C.R.L. for hospitality during
the period of his stay in Tokyo during which this paper was written.
\end{acknowledgements}


\begin{thebibliography}{999}
\bibitem{} Babul A., White S.D.M., 1991, MNRAS 253, 31P 
\bibitem{} Bardeen J.M., Bond J.R., Kaiser N., Szalay A.S., 1986, ApJ  304, 15 
\bibitem{} Barrow J.D., Silk J., 1981, ApJ 250, 432
\bibitem{} Bartlett J.G., Silk J., 1993, ApJ 407, L45
\bibitem{} Borgani S., 1990, A\&A 240, 223
\bibitem{} Bower R.G., Coles P., Frenk C.S., White S.D.M., 1993, ApJ  405, 403 
\bibitem{} Cen R. Ostriker J. P., 1992, ApJ 399, L113
\bibitem{} Cen R.Y., Gnedin N.Y., Kofman L.A., Ostriker, J.P., 1992 preprint
\bibitem{} Coles P., 1993, MNRAS, 262, 1065
\bibitem{} Davis M., Peebles P.J.E., 1977, ApJS., 34, 425
\bibitem{} Del Popolo A., Gambera M., 1998a, A\&A 337, 96
\bibitem{} Del Popolo A., Gambera M., 1998b, A\&A in print (see also SISSA preprint, astro-ph/9806044)
\bibitem{} Del Popolo A., Gambera M., 1998c, Proceedings of the VIII Conference on Theoretical Physics: General Relativity and Gravitation - Bistritza -  June 15-18, 1998 - Rumania
\bibitem{} Efstathiou G., Sutherland W.J., Maddox,S.J., 1990a, Nat., 348, 705 
\bibitem{} Efstathiou G., Kaiser N., Saunders W. et al. 1990b, MNRAS 247, 10p 
\bibitem{} Governato F., Babul, A., Quinn T., Tozzi P., Baugh C. M., Katz N., Lake G., 1998, SISSA preprint astro-ph/9810189
\bibitem{} Groth, E.J., Peebles P.J.E., 1977, ApJ 217, 385 
\bibitem{} Hoffman, Y., 1986, ApJ 301, 65
\bibitem{} Holtzman J., 1989, ApJS 71, 1
\bibitem{} Holtzman J., Primack J., 1993, Phys. Rev. D43, 3155 
\bibitem{} Kaiser, N., 1984, ApJ  284, L9
\bibitem{} Kauffmann G., Colberg J.M., Diaferio A., White S.D.M., 1998, SISSA preprint astro-ph/9805283
\bibitem{} Lacey C., Cole S., 1994, MNRAS 271, 676
\bibitem{} Limber, D.N., 1954, ApJ  119, 655
\bibitem{} Loveday J., Efstathiou G., Peterson B.A., Maddox S.J., 1992, ApJ 400, L43
\bibitem{} Maddox S.J., Efstathiou G., .Sutherland W.J., Loveday J., 1990, MNRAS 242, 43p
\bibitem{} Maddox S.J., Efstathiou G., Sutherland W.J., 1996, MNRAS 283, 1227
\bibitem{} Mo H.J., White S.D.M., 1996, MNRAS 282, 347
\bibitem{} Monaco P., 1995, ApJ 447, 23 
\bibitem{} Ostriker J.P., 1993, ARA\&A. 1993, 31, 689
\bibitem{} Peacock J.A., 1991, MNRAS, 253, 1p
\bibitem{} Peacock J.A., 1998, SISSA preprint astro-ph/9805208
\bibitem{} Peacock J.A., Nicholson D., 1991, MNRAS, 253, 307 
\bibitem{} Peebles P.J.E., 1980, The large scale structure of the Universe, Princeton university press 
\bibitem{} Peebles P.J.E., 1984, ApJ 284, 439
\bibitem{} Peebles P.J.E., 1990, ApJ 365, 27 
\bibitem{} Press W.H., Schechter P., 1974, ApJ 187, 425 
\bibitem{} Ryden B.S., 1988, ApJ 329, 589 
\bibitem{} Saunders, W., Frenk C., Rowan-Robinson M. et al., 1991, Nature 349, 32
\bibitem{} Schaefer R.K., 1991, Int. J. Mod. Phys. A6, 2075
\bibitem{} Schaefer R.K., Shafi Q., 1993, Phys. Rev., D47, 1333 
\bibitem{} Schaefer R.K., Shafi Q., Stecker F., 1989, ApJ 347, 575 
\bibitem{} Shafi Q., Stecker F.W., 1984, Phys. Rev. D29, 187
\bibitem{} Szalay A.S., Silk J., 1983, ApJ 264, L31 
\bibitem{} Turner M.S., 1991, Phys. Scr. 36, 167 
\bibitem{} Valdarnini R., Bonometto S.A., 1985, A\&A 146, 235
\bibitem{} Villumsen J.V., Davis M., 1986, ApJ 308. 499
\bibitem{} Vogeley M.S., Park C., Geller M.J., Huchra J.P., 1992, ApJ 391, L5
\bibitem{} Willmer C.N.A., Nicolaci da Costa L., Pellegrini P.S., 1998, SISSA preprint astro-ph/9803118

\end{thebibliography}
\end{document}